# Preparation of light-atom tips for Scanning Probe Microscopy by explosive delamination

T. Hofmann, J. Welker, F.J. Giessibl

University of Regensburg, Institute of Experimental and Applied Physics II, Experimental Nanoscience, Universitätsstraße 31, 93053 Regensburg, Germany

**Abstract.** To obtain maximal resolution in STM and AFM, the size of the protruding tip orbital has to be minimized. Beryllium as tip material is a promising candidate for enhanced resolution because a beryllium atom has just four electrons, leading to a small covalent radius of only 96 pm. Besides that, beryllium is conductive and has a high elastic modulus, which is a necessity for a stable tip apex. However beryllium tips that are prepared ex situ, are covered with a robust oxide layer, which cannot be removed by just heating the tip. Here we present a successful preparation method that combines the heating of the tip by field emission and a mild collision with a clean metal plate. That method yields a clean, oxide-free tip surface as proven by a work function of $\Phi_{exp} = 5.5$ eV as deduced from a current-distance curve. Additionally, a STM image of the Si-(111)-(7x7) is presented to prove the single-atom termination of the beryllium tip.

Over the last decade the resolution in non contact atomic force microscopy (NC-AFM) has been increased tremendously. Recently, the chemical structure of a molecule could be resolved by terminating the tip with a CO molecule [10]. In 2004, the orbital structure of the tungsten atom was imaged by scanning a graphite surface with a tungsten tip [12]. In both cases the small covalent radii of the probing atoms, namely oxygen ($R_{cov} = 66$ pm) [10, 4] and carbon ($R_{cov} = 76$ pm) [12, 4], account for the high resolution. For combined STM/AFM, it is desirable to use a light metal with comparably small covalent radius as tip material. We chose beryllium, because it has the smallest covalent radius of all metals with $R_{cov} = 96$ pm [4]. Another benefit of beryllium is its high binding energy compared to other light metals [11], ensuring a stable tip apex. The surface of beryllium is covered by a native oxide layer consisting of beryllium oxide with a thickness of around 1 nm [16].

The removal of the oxide layer is challenging because it has a higher boiling and melting point than pure beryllium. However, the removal of the oxide layer is mandatory due to its high specific resistance ($\rho_{BeO} > 10^{15}$ Ωcm [3]). The most common method, used for preparation of diverse scanning probe tips, is heating the tip apex until adsorbates and the oxide layer are evaporated. Tips can be heated, for example, by driving a direct current through the tip [17], bombard the tip with electrons [7], directly touch a hot filament or resistively heat the tip by field emission [12]. Due to our sensor setup only electron bombardment and field emission can be used because the glue, which fixes the tip to the prong of the tuning fork [8], withstands only temperatures up to 150 °C. Other techniques can lead to a higher temperature at the tip shaft. Unfortunately it is not possible to successfully prepare the beryllium tips by just heating them with these methods.

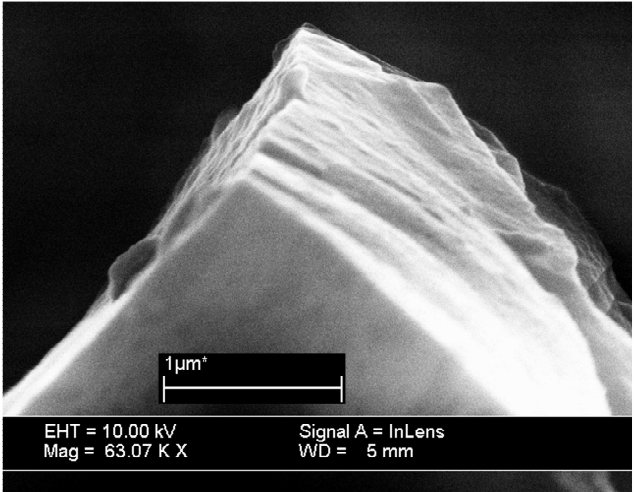

**Fig. 1.** SEM image of a cleaved beryllium fragment used as scanning probe tip. On the macroscopically flat cleavage planes many small protrusions are visible.

In this paper, we present an effective method for preparing the beryllium tips in UHV. We heat the tips by field emission, while carefully touching a clean metal plate. With this procedure it is possible to get a clean, oxide-free tip surface, as the obtained STM image (Fig. 3) and the current-distance curve (Fig. 4) show.

The qPlus [8] is characterized by a very stiff cantilever allowing oscillation amplitudes as small as a few hundred picometers. Therefore tips with various shapes can be mounted to the prong without reducing the performance due to long range van der Waals forces [9]. We used beryllium tips produced by crushing a polycrystalline beryllium lump (purity of 99.5%) with a gripper. To meet safety regulations the procedure is conducted in a liquid environment. Possible tips are selected from the resulting fragments under an optical microscope. Most of the fragments show macroscopically flat cleavage planes due to the brittle fracture behaviour of beryllium. In the SEM image in figure 1 the front region of a selected beryllium piece is shown. It indicates that the surfaces of the cleavage planes exhibit many small protrusions. The selected beryllium fragments are glued conductively onto a gold wire, which is attached to the front side of the free prong of the tuning fork with non conductive glue. The other end of the gold wire is connected to the terminal for the bias voltage. With this setup both electrodes of the tuning fork are electrically isolated from the bias voltage.

The sensor is transferred to a room temperature UHV chamber (base pressure $5 \times 10^{-11}$ mbar). To achieve stable tunneling operation it needs to be cleaned. The reason is that, besides adsorbates, the native oxide layer, consisting of highly insulating beryllium oxide, needs to be removed. One possibility is to locally heat the tip by a field emission current. As shown in figure 2, a negative voltage of up to 1 kV is applied between the tip (cathode) and a grounded metal plate (anode), e.g. made of copper (used for trial experiments) or tantalum. Due to the high electric field the width of the surface potential barrier at the tip is reduced and electrons can tunnel into the vacuum gap. These are then accelerated to the anode resulting in a field emission current, which is described by the Fowler-Nordheim equation [6]:

$$I_{FE} = C_1 \frac{(E_F / \Phi)^{1/2}}{E_F + \Phi} F^2 \exp(-C_2 \frac{\Phi^{3/2}}{F}), \quad (1)$$

where $E_F$ is the fermi energy, $\Phi$ the work function and $F$ the electric field; $C_1$ and $C_2$ are constants. The cathode material, thus the tip, is resistively heated by the field emission current $I_{FE}$. As the field emission current depends exponentially on the electric field, according to equation 1, only small protrusions, enhancing the electric field, carry a high field emission current. Therefore just these spots on the surface are heated. The surface of the cleaved beryllium tips is macroscopically smooth exhibiting many small protrusions (see figure 1). This leads to a complex electric field between tip and metal plate. Therefore the distance between tip and metal plate has to be a few micrometers to get an emission current of a few microampere at a tip bias of 1 kV. This current is not sufficient to heat the tip beyond the melting temperature of beryllium. To increase the current the tip-plate distance has to be reduced.

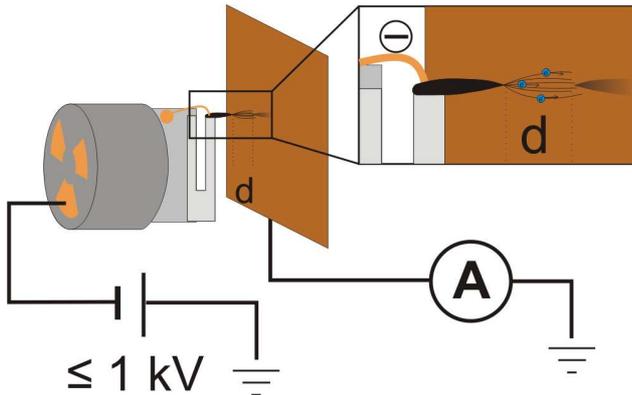

**Fig. 2.** Scheme of the setup during field emission. A negative voltage of up to 1 kV is applied to the tip. By controlling the distance $d$ the electric field and therefore the field emission can be controlled. The field emission current is limited to 1 mA to prevent excessive heating of the tip.

For the preparation of the beryllium tips it is not enough to just heat the tip apex, as the oxide layer has a higher melting ($T_M \approx 2600\,°C$) and boiling point ($T_B \approx 3200\,°C$) than the covered beryllium ($T_M \approx 1300\,°C$). In addition, the oxide layer is very rigid. Therefore it is necessary to melt the tip apex and additionally, break the oxide mechanically. For this purpose, the Beryllium tip is brought into contact with the metal plate while a voltage of 1 kV is applied. Shortly before mechanical contact visible arcing between tip and sample occurs. The spark is initiated by the field emission current. A possible reason for the arc is that the heat production at the tip apex is high enough to evaporate tip material. In the vacuum gap the evaporated atoms from the cathode are ionized and form an arc [5]. The current needed to heat the tip apex to evaporation temperature is just reached at tip-plate distances below 1 um. At these distances the forces on the tip are already high enough that the tip snaps to the metal plate. At contact of the tip to the metal plate the oxide breaks open and the tip is melted to the plate. By retracting the tip for a few micrometers it is torn off the metal plate and again a vacuum arc due to high field emission current is generated. Optical examination of an anode made of copper showed grey, glossy dots where the tip touched the plate. Although the melting point of copper ($T_M^{Cu} \approx 1100\,°C$) is lower than the one of beryllium ($T_M^{Be} \approx 1300\,°C$), the grey spots on the copper surface indicate that the beryllium is melted to the copper plate. Therefore it is believed that a contamination of the beryllium surface with copper is unlikely. To further minimize the probability of contaminating the beryllium surface with the anode material, the copper plate was exchanged with a tantalum plate after the first experiments. As tantalum has a very high melting point ($T_M^{Ta} \approx 3000\,°C$), the probability that anode material is melted and transferred to the beryllium tip is minimized. During the preparation no increase in pressure can be observed. It can be assumed, that the released material is ionized and implanted into the anode. Therefore the tip can be prepared after the sample without contaminating the prepared sample surface.

First measurements with the beryllium tips, prepared by the combination of partially melting the tip by high field emission current and a mild collision with the anode, are conducted on the Si-(111)-(7x7) surface. The sample is prepared by the common method of flashing the sample several times to a temperature of over $1200\,°C$, quickly cooling down to $950\,°C$ and slowly cooling down to $700\,°C$. To avoid a contamination of the tip during scanning, the tunneling current is set to a small value of 50 pA at a bias voltage of $V = -2\,V$. Therefore the chance of touching the surface while scanning a large sample area with many steps is minimized.

After the tip preparation and the approach, a stable tunneling condition is observed and it is possible to immediately obtain steps at the maximal scan size of $560 \times 560\,nm^2$. After reducing the size of the scan area, images with atomic resolution can also be obtained.

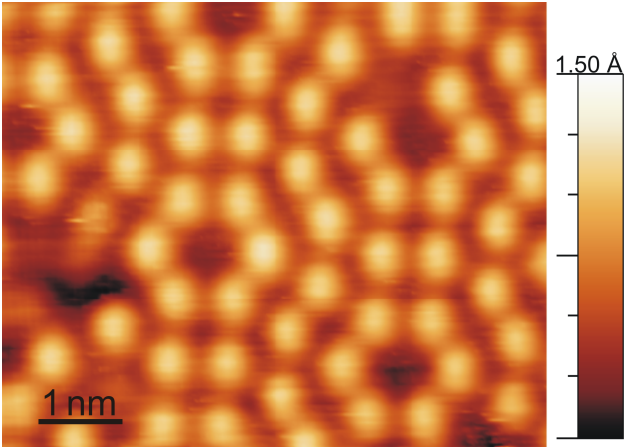

**Fig. 3.** STM image of the Si-(111)-(7x7) surface recorded with a beryllium tip after tip preparation. Image size: $5.4 \times 6.5$ nm$^2$. Imaging parameters: $V = -1.5$ V and $I_t = 69$ pA.

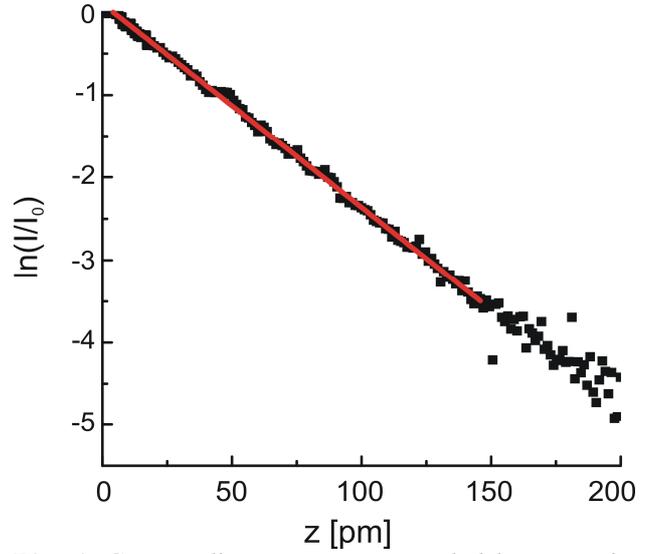

**Fig. 4.** Current-distance curve recorded by retraction of a beryllium tip from the Si-(111)-(7x7) surface. Initial parameters: $V = -0.5$ V and $I_0 = 2$ nA. The fitted line shows a slope of $0.024$ pm$^{-1}$ corresponding to a decay constant of $\kappa = 1.2$ Å$^{-1}$.

The image in figure 3 is recorded with a bias voltage of $V = -1.5$ V and a tunneling current of $I_t = 69$ pA and shows the (7x7) surface reconstruction of the silicon surface with some defects.

For showing the metallic character of the tip, a current-distance curve, shown in figure 4, is recorded on the Si-(111)-(7x7) surface at a bias voltage of $V = -0.5$ V and a maximal tunneling current of $I_t = 2$ nA. The curve exhibits an exponential dependence of the current on the distance with a decay constant $\kappa$ of about $1.2$ Å$^{-1}$. The work function $\Phi_{exp}$ is calculated to a value of $\Phi_{exp} = \kappa^2 \hbar^2 / 2m_e = 5.5$ eV. With a value of $\Phi_{Si} = 4.7$ eV for the silicon surface [1] the work function of the tip is calculated to $2\Phi_{exp} - \Phi_{Si} = 6.3$ eV. As the work function of Beryllium $\Phi_{Be} = 5.0$ eV [13] is comparable to other metals like copper $\Phi_{Cu} = 4.7$ eV or Tantalum $\Phi_{Ta} = 4.3$ eV [14] and is $1.3$ eV smaller than the calculated value, the current-distance measurement can just prove that the tip surface is metallic. A remaining oxide layer would decrease the work function to a value well below $5.0$ eV [2]. The current-distance curve therefore shows that the oxide can be removed successfully by the described method.

The Beryllium surface is very reactive. Therefore one monolayer of beryllium oxide is formed again after a few hours under ultrahigh vacuum conditions (base pressure below $3 \times 10^{-11}$ mbar) [15]. Due to the high resistance of the oxide it is impossible to get a stable tunneling current using a tip contaminated with beryllium oxide. As a result it is only possible to measure with a beryllium tip for a few hours, before it has to be prepared again.

In summary, we have shown that a combination of melting the tip apex by a high field emission current and a mild collision with a metal plate proves to be a successful preparation method for Beryllium tips. Immediately after the approach a STM image with atomic resolution of the Si-(111)-(7x7) surface can be obtained. Additionally, a current-distance curve is recorded indi-cating a work function of $\Phi_{exp} = 5.5\,\text{eV}$ being in good agreement with the literature value of $\Phi_{Be} = 5.0\,\text{eV}$. Because beryllium is very reactive, the tip has to be prepared again after a few hours at room temperature. However, we expect that the life time of the tip will be at least weeks in a helium-temperature UHV environment.

**Acknowledgment.** We wish to thank Veeco Instruments Inc., Plainview, for providing us the UHV equipment and B. Birkner for the SEM image of the beryllium fragment.